\documentclass[preprint,12pt, a4paper]{elsarticle}

\usepackage{amssymb}
\usepackage{hyperref}

\journal{Software Impacts}

\DeclareRobustCommand{\osquare}{O\textsuperscript{2}}

\begin{document}

\begin{frontmatter}



\title{FieldView: An interactive software tool for exploration of three-dimensional vector fields}


\author{Piotr Nowakowski\corref{cor1}}
\ead{piotr.nowakowski.dokt@pw.edu.pl}

\author{Przemysław Rokita}
\ead{P.Rokita@ii.pw.edu.pl}

\cortext[cor1]{Corresponding author}

\address{Faculty of Electronics and Information Technology, Warsaw University of Technology, Nowowiejska 15/19, 00-665 Warsaw, Poland}

\begin{abstract}
Vector fields are one of the fundamental parts of mathematics which are key for modelling many physics phenomena such as electromagnetic fields or fluid and gas flows. Fields found in nature often exhibit complex structures which can be difficult to analyze in full detail without a graphical representation. In this paper we introduce FieldView --- an application targeted at academics and industry professionals wanting to visualize vector fields using various rendering techniques. FieldView uses a novel way of evaluation of the field and construction of the displayed elements directly on the GPU which is based on the recently introduced mesh shader graphics pipeline by NVIDIA. Research done for FieldView development has found practical application in the form of magnetic field visualisation integrated in the data processing framework developed by ALICE experiment at CERN.

\end{abstract}


\end{frontmatter}


\section{Motivation}

3D vector field visualisation is a mature topic with broad applications in medical, machine engineering, physics and weather sciences~\cite{techniques_overview}. The exact technique utilized usually
depends on the characteristics of the field itself, as all have their advantages and weaknesses when used on different input data. Vector field visualisation is generally divided into four groups: direct, dense texture-based, feature-based and geometric. Direct methods consist of placing some kind of glyph (e.g. an arrow) at each sample point in the evaluated domain and are computationally inexpensive~\cite{vector_plots,truflow,local_icons}. Their disadvantages include occlusion problems and visual complexity for large vector fields, limiting their usefulness. Texture-based techniques (such as Line Integral Convolution) use an image, usually a some kind of noise, which is then smeared or stretched according to properties of the vector field~\cite{spot_noise,auflic,texture_advection}. They tend to highlight interesting features of fields, such as sources, sinks and vortexes and provide a lot of detail; however, they suffer similar occlusion problems as the direct methods and are better suited in 2D environments. Feature-based techniques are similar to the direct methods in that certain glyphs are displayed; however, they are not representing the field itself but some selected set of its features, such as already mentioned sources, sinks or vortexes~\cite{galilean,saddle_connector,vortices}. This significantly reduces the amount of visual information displayed and as such is best suited for fields with a large number of mentioned elements. These techniques require more computational power as the field data usually needs to be pre-processed to find such features. Finally, in geometric methods, trajectories are computed from a set of seed points which are either animated (hence becoming particle-based visualisations~\cite{autonomous_particles}), displayed directly (streamlines)~\cite{streamlines} or used to construct other geometric shapes, such as tubes or ribbons~\cite{streamribbons,tubes_ray}.

One of the key research directions in vector field visualisation is development of algorithms dedicated for graphics cards. With constant advancements in architecture and performance of GPUs more
and more visualisation techniques can be performed in real time.

In this paper we present FieldView --- an ongoing research project which explores the capabilities of the mesh shader pipeline~\cite{mesh_shaders_turing_web} of modern graphics cards (introduced by NVIDIA with the Turing micro-architecture~\cite{nvidia_turing_2018, turing}) to visualise vector fields using geometric techniques. To our best knowledge FieldView is one of the first scientific projects to use the mesh shader pipeline~\cite{mesh_lod,mesh_terrain,meshlet_bounds} to visualise 3D vector fields.

Unlike a classical approach, where the main computer processor is used to evaluate the vector field and generate lists of vertices, our method focuses on moving all relevant calculations to the GPU itself. This offers much more flexibility in terms of visualisation possibilities and reduces the amount of data needed to be exchanged between the main memory and the GPU. To make the application easy to use for everyone we have developed a user-friendly graphical user interface. Within the GUI the visualisation parameters, such as the field equation and selection of display technique, can be tweaked.

FieldView core rendering functionality has been also used by our team to enhance Event Display, a particle collision visualisation tool which is part of \osquare{}, the data processing framework developed by ALICE experiment at CERN~\cite{otwodesign}.

The rest of the paper is organized as follows. In Section~\ref{sec:arch} the software architecture is presented with special attention put on the novel mesh shader graphics pipeline which allowed this tool to be created. Examples showing all features of FieldView with screenshots are described in Section~\ref{sec:examples}. Impact and possible further improvements are discussed in Section~\ref{sec:impact}. Final conclusions are presented in Section~\ref{sec:conclusions}.

\section{Software architecture}
\label{sec:arch}

\begin{figure}
    \centering
    \includegraphics[width=\linewidth]{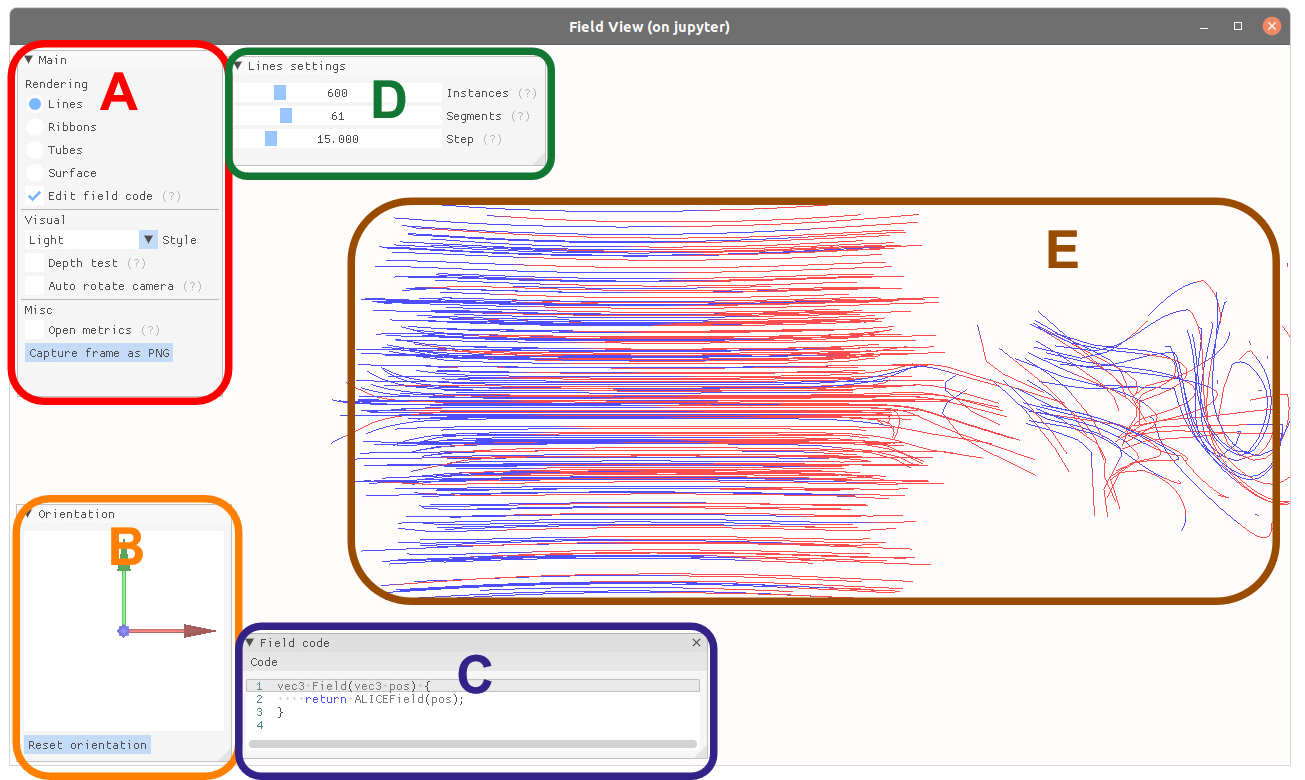}
    \caption{Main window of the application.}
    \label{fig:menu}
\end{figure}

FieldView is written in C++ language and uses features of the C++17 standard. Access to the OpenGL 4.6 API~\cite{opengl}, used for rendering, is provided either by the GLFW~\cite{glfw} or EGL~\cite{egl} library chosen at compile time (the latter being used for building the application for work in an non-interactive way). Internally the software is constructed from a set of small static libraries: \texttt{common}, which handles window/rendering context creation; \texttt{viz}, which handles magnetic field data loading (see Section~\ref{sec:examples}), screenshots and general options; and \texttt{viz\_instance}, which holds implementations of the actual field visualisation techniques. \textit{Dear ImGui}~\cite{imgui} library was used to implement GUI elements of the application (see Figure~\ref{fig:menu} for reference on how the main window is presented).

As mentioned in the introduction FieldView uses relatively new technology, which is the mesh shader pipeline, to render graphics. It is made of only three programmable stages: the task shader, the mesh shader and the fragment shader. The task shader is responsible for launching instances of the mesh shader and as such is capable of dynamically adjusting the amount of work scheduled to be done by the GPU based on e.g. distance of a rendered object from the camera. The mesh shader is responsible for generating vertex data which is then used by the hardware rasterization stage. The fragment shader performs the same task as in the standard pipeline (calculating color of displayed pixels). Task and mesh shaders, like compute shaders, are based on a collaborative multi-threaded design. This means that each shader invocation launches a predefined number of local GPU threads that operate on shared memory. The new pipeline allows for certain tasks, such as geometry generation on the GPU, to be performed more efficiently than in the standard pipeline (where a geometry shader needs to be used to achieve similar results).

FieldView is licensed under the GNU General Public License v3.0~\cite{gnugpl} and is compatible with GNU/Linux and Windows. An NVIDIA RTX graphics card is required to run the software due to its reliance on the mesh shader pipeline, which right now is only available in this family of video cards. We hope that this functionality will be soon standardized in the OpenGL API and offered by other GPU vendors as well.

\section{Illustrative examples}
\label{sec:examples}

\begin{figure}
    \centering
    \includegraphics[width=\linewidth]{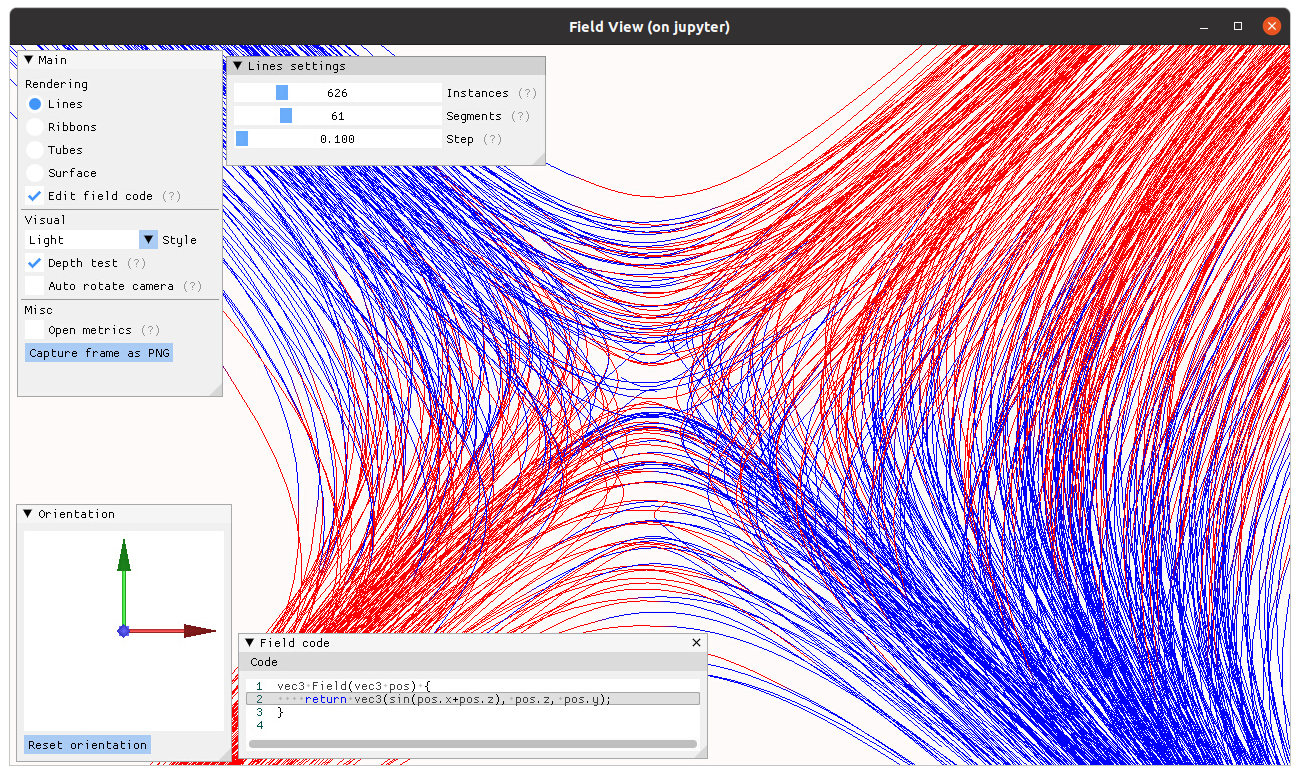}
    \caption{FieldView with selected line renderer, displaying field \(F = (\sin(x+z), z, y) \).}
    \label{fig:lines}
\end{figure}

\begin{figure}
    \centering
    \includegraphics[width=\linewidth]{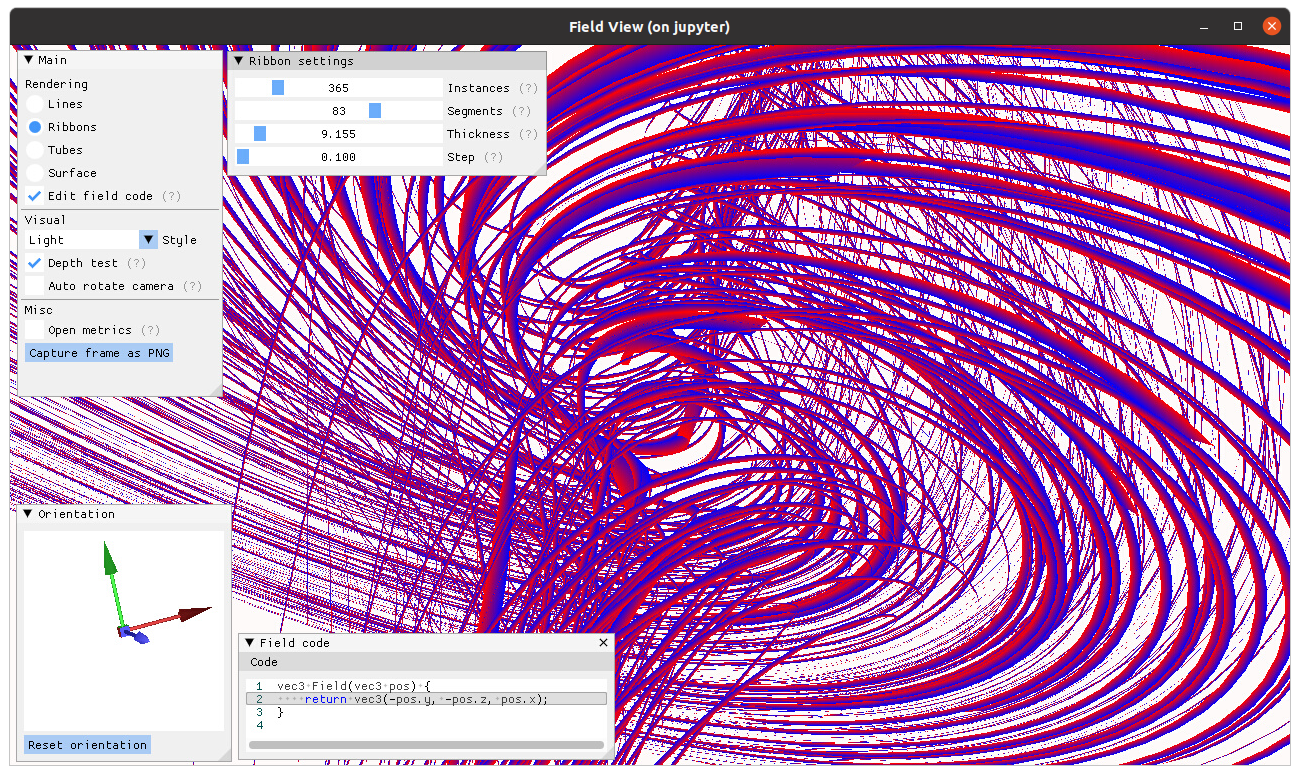}
    \caption{FieldView with selected ribbon renderer, displaying field \(F = (-y, -z, x) \).}
    \label{fig:ribbons}
\end{figure}

\begin{figure}
    \centering
    \includegraphics[width=\linewidth]{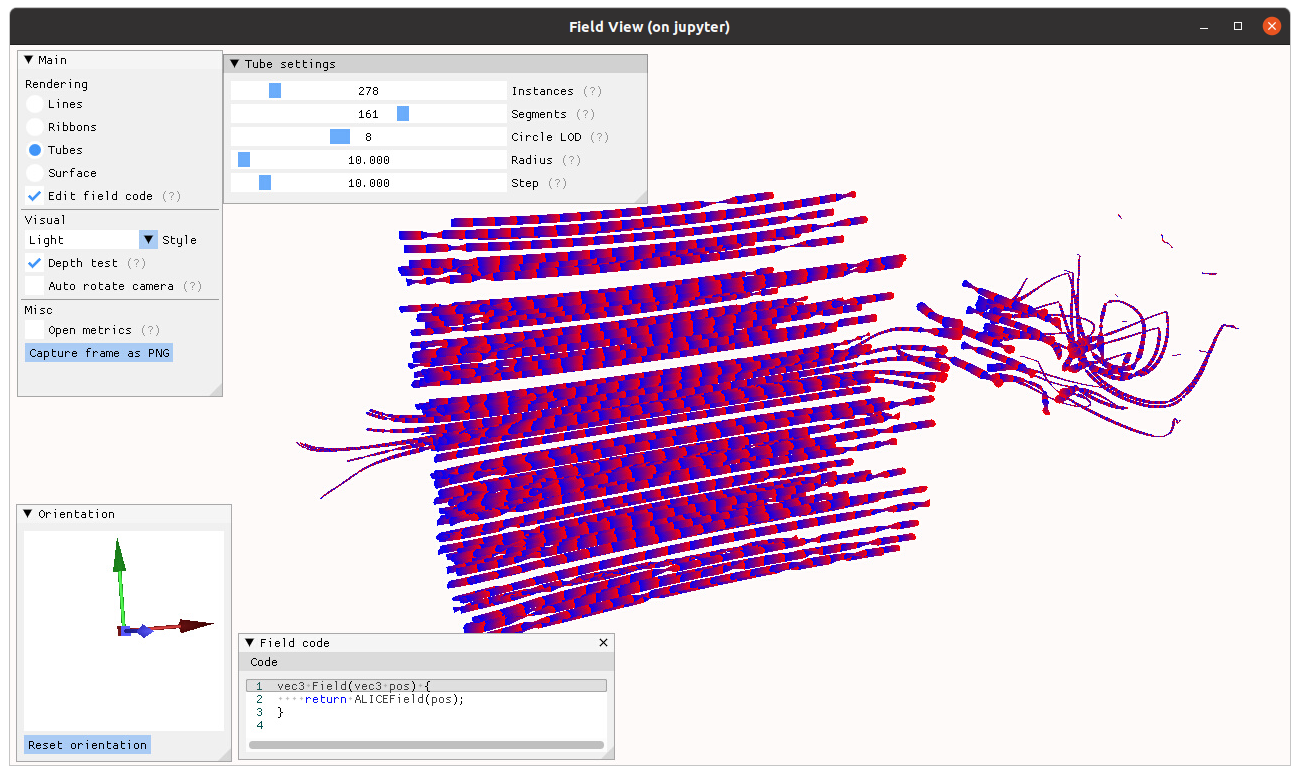}
    \caption{FieldView with selected tube renderer, displaying ALICE detector field.}
    \label{fig:tubes}
\end{figure}

\begin{figure}
    \centering
    \includegraphics[width=\linewidth]{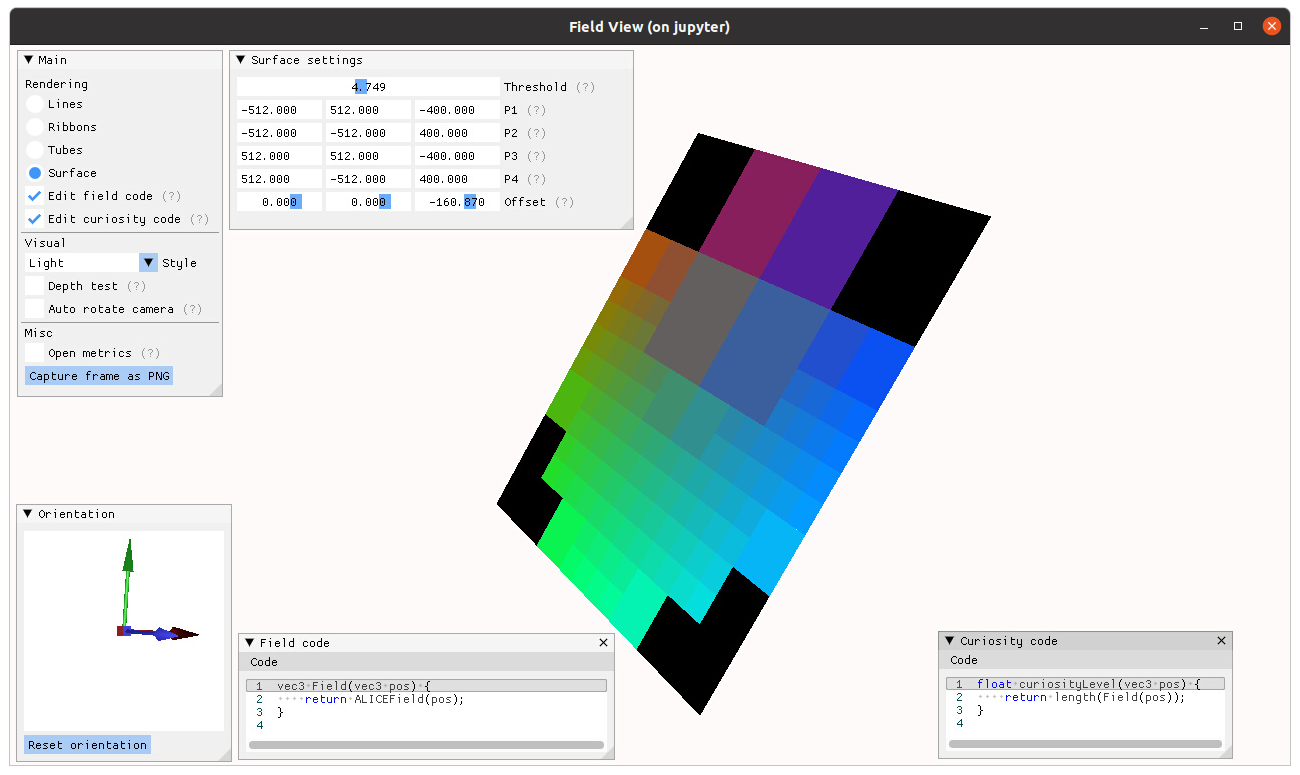}
    \caption{FieldView with selected surface renderer, displaying ALICE detector field.}
    \label{fig:surface}
\end{figure}

FieldView is a GUI application which enables the user to explore the graphical representation of 3D vector fields using various rendering techniques. In the current implementation, the following four methods are available:
\begin{itemize}
    \item Lines --- representation of the field's stream lines using single-pixel-thick curves. The simplicity of this approach allows for decent clarity even with high density of primitives at a cost of displaying only the general direction of the field (Figure~\ref{fig:lines}).
    \item Ribbons --- representation of two nearby stream lines connected into a ribbon of variable width. This approach is capable of showing twists, sources and sinks of the field in addition to the direction (Figure~\ref{fig:ribbons}).
    \item Tubes --- representation of the field's stream lines using cylindrical sections of variable radius connected together into tubes. This approach is capable of displaying the strength of the field in an intuitive way (by adjusting the radius of the tubes accordingly) in addition to all properties offered by the ribbon approach (Figure~\ref{fig:tubes}).
    \item Surface --- a method designed to be the most interactive; it is based off of a configurable planar cross section of the field. The field is sampled over the plane and then processed by a \texttt{curiosityLevel()} function which returns a floating-point value indicating how interesting a particular area of the plane is to the user. Average value of curiosity of nearby samples is tested by the algorithm against an adjustable threshold and averaged if it falls below it. The \texttt{curiosityLevel()} is fully user-defined, allowing a person using the application to enhance the desired characteristic of the field (Figure~\ref{fig:surface}).
\end{itemize}

When the application is first started, the following window is displayed --- see Figure~\ref{fig:menu}. Section A (titled \textit{Main}) controls the general aspects of rendering. Here the user can select one of the visualisation methods (Lines, Ribbons, Tubes, Surface) as well as open the field equation editor (which is visible as Section C). Settings which control visual aspects are placed underneath. First is a style selection (light or dark theme), which allows users to adjust the appearance of the application to their current environment (time of day, screen/projector use etc.). Following that there is a checkbox to enable depth testing (which causes the field primitives to be correctly displayed in terms of occlusion) and an option to automatically rotate the view. Finally, an option to show diagnostic information and a button to capture a screenshot is presented.

Section B, titled \textit{Orientation}, gives users the option to manually adjust the camera position for the field visualisation. This is done by clicking and dragging the mouse in the the square area. An indicator of the camera's current position, relative to the default, is displayed in the form of three arrows representing axis of the coordinate system (x, y, z). Underneath this area there is a button which allows quick reset of the camera position to the default.

Section C is the equation editor. Here users can input their own mathematical formula to visualise the desired field. What is written in the editor is directly injected into a shader directly responsible for construction of geometry, written in GLSL. As such, any built-in GLSL math function can be used by the user in the field equation (trigonometric, cross and dot products, min and max, clamps, exponents etc.). By default (to showcase the capabilities of the application) the magnetic field of the ALICE~\cite{nowakowski_distributed} collision detector from CERN is used. To apply any modifications done to the code, users have to select \textit{Code} -- \textit{Save \& Apply} from the menu. If there is a syntax error in what was written, the editor will show an error message. To reset the equation back to the default, the option \textit{Code} -- \textit{Reset \& Apply} should be chosen from the menu.

Section D contains settings which directly control the chosen visualisation method. As such the number and type of available settings change depending on the the method which was selected in Section A. In general, users can adjust the number of rendered shapes, the number of generated segments inside them and their thickness.

Section E is the main render area where the field visualisation is displayed.

\section{Impact}
\label{sec:impact}

As already mentioned in the introduction, FieldView is one of the first few scientific projects which use the mesh shader pipeline~\cite{mesh_lod,mesh_terrain,meshlet_bounds} to visualise 3D vector fields and can serve as a reference on how to integrate this technology in other fields of study. The modular design of the application ensures that other visualisation techniques can be easily added in the future.

Parts of the program, in the form of ALICE detector magnetic field visualisation, have been used in our recent accepted code submission for Event Display, a particle collision visualisation application developed by the ALICE Collaboration at CERN~\cite{alice_o2}. The Event Display will use it as a visual guide for detector operators to quickly asses the correct state of both hardware and software components of ALICE experiment during periods of data taking.

Lastly, FieldView is under consideration for inclusion in the International ALICE Masterclass project~\cite{alicemasterclass} organized by the International Particle Physics Outreach Group~\cite{ippog}. The aim of the Masterclass project is education, popularization of physics and mission of CERN among students. FieldView could be used there as part of the introductory lecture as a visual aid during presentation of ALICE detector hardware capabilities.

\section{Conclusions}
\label{sec:conclusions}

The paper describes FieldView --- a graphical user interface application designed for interactive visualisation of 3D vector fields. The current version of the tool implements four techniques of geometrical representation of the field (lines, ribbons, tubes, surface), although more options can be added in the future due to modular design of the renderer code. The application uses a novel technology (the mesh shader pipeline) to render the displayed images, which is not yet fully explored in the currently available body of scientific work. 

The program offers an example of a complex field built-in in the form of magnetic field data from the ALICE particle detector, but any other field equation can be provided by the user and immediately visualised thanks to the equation editor available in the GUI. FieldView can be used by students and lecturers as a visual aid during study of the topic of vector fields.

\section*{Declaration of competing interest}

The authors declare that they have no known competing financial interests or personal relationships that could have appeared to influence the work reported in this paper.

\section*{Acknowledgements}

This work was supported by the Polish National Science Centre under agreements no. UMO-2016/22/M/ST2/00176, UMO-2016/21/D/ST6/01946, no. UMO-2017/27/B/ST2/01947,  no. 2021/43/D/ST2/02214, by the Polish Ministry for Education and Science under agreements no.  2022/WK/01 and 5236/CERN/2022/0, as well as by the IDUB-POB-FWEiTE-1 project granted by  Warsaw University of Technology under the program Excellence Initiative: Research University (ID-UB).

\bibliographystyle{elsarticle-num}
\bibliography{biblio}

\end{document}